# Going Green: A Holistic Approach to Transform Business

**Dr. Sajal Kabiraj***

Professor, FCRIMS, University of Mumbai, India

**Dr. Vinay Topkar**

Professor, VJTI, University of Mumbai, India

**R.C. Walke**

Doctoral Ph.D Student, VJTI, University of Mumbai, India

*Abstract:*
In recent years environmental and energy conservation issues have taken the central theme in the global business arena. The reality of rising energy cost and their impact on international affairs coupled with the different kinds of environmental issues has shifted the social and economic consciousness of the business community. Hence, the business community is now in search of an eco-friendly business model. This paper highlights the concept of green business and their needs in the current global scenario.

1. Green Computing: A Gate way to Green Business

Green computing is the study and practice of efficient and eco-friendly computing resources. Now it is not only under the attention of environmental organizations, but also other business industries. In recent years, companies in the computer industry have come to realize that going green inspired business model is the best for their interest. In other words, the green computing is nothing but the efficient use of computers and computing methods. Recently, the triple bottom line (abbreviated as "TBL" or "3BL" and also known as "people, planet, profit") is becoming a new measure for business performance (Brown, *et. al*, 2006). These three pillars capture an expanded spectrum of values and criteria for measurement of organizational (and societal) success mainly the economical, ecological and social status of the organization. This triple bottom line concept is important when it comes to anything green and the same goes for green computing. Many businesses simply focus on a bottom line, rather than a green triple bottom line particularly in computer industry. The idea of green business is to make the whole process surrounding the computers friendlier to the environment, economy, and society. This means manufacturers create computers in such a way that reflects the triple bottom line positively. Once computers are sold the business or people use the computer in a green way by reducing power usage and disposing of them properly or recycling them efficiently. This idea is to make computers from beginning to end a green product one.

In 2009, Technology Business Research (TBR) announced that Dell took the No. 1 position in its inaugural Corporate Sustainability Index (CSI) Benchmark Report. The report measures the environmental initiatives of 40 companies in the computer hardware, software, professional services & network and telecommunication sectors. Scoring 317.9 points, Dell is 52 points ahead of the firm positioned second in the overall CIS ranking index (Round Rock, 2009). IBM also recently launched consulting services, which is based on the Lean Six Sigma principles of efficiency. This consulting service is aimed at examining the use of energy and water and subsequently providing the control measures to conserve the energy. According to the company sources, IBM in 1990 saved around 4.6 billion Kwh of electricity and prevented emission of almost 3 million metric tons of carbon dioxide. So, essentially a reduction in wastage and recycling of the used materials is required to ensure the green IT. There have been many approaches to green computing suggested by Mike (Mike Ebbers *et al*, 2009). According to VMware Inc. report the global leader in virtualization solutions from the desktop to the data center announced the opening of a new green IT data center in East Wenatchee, Washington. Throughout its design and build out, VMware chose industry best practices to create an energy-efficient facility that utilizes cutting-edge technology and maximizes the use of VMware virtualization software. As a result, VMware has achieved $5 million savings per year from the said data centre (Rakesh Kumar, 2007).

According to SUN, Today's modern network economics requires high computing capability for searches, Web services, e-commerce, traffic control, or supply chain management. The said services required high power computing with significant capacity. SUN approached the problem by pushing the physical technology of the CPU frequency namely, the number of cycles that a piece of silicon can do. But it quickly run into the law of physics which states that when the transistors switch-on and switched-off quickly, then there is a corresponding amount of power consumption and the heat generated by the transistors grows proportionately. Considering that networking it has a quadratic effect where the interaction grows geometrically. The industry has reached a point where it has driven power consumption of these products to a point that deviates from what the customer can utilize (MTB, 2009). One of the VIA Technologies' ideas is to reduce the "carbon footprint" of users i.e the amount of greenhouse gases produced, which is measured in units of carbon dioxide. Greenhouse gases naturally blanket the Earth and are responsible for its more or less stable temperature. An increase in the concentration of the main greenhouse gases like carbon dioxide, methane, nitrous oxide, and fluorocarbons are believed to be responsible for Earth's increasing temperature. This change could lead to severe floods, droughts, rising sea levels, and other environmental effects, affecting both life and the world's economy (Greg Horn, 2006). The energy star program encourages manufactures to create energy-efficient devices that require little power in idle condition. For example, many devices switch to standby mode after a specified number of inactive minutes. Personal computers, monitors and printers should comply with the energy star program, which was developed by the United States Department of Energy (DOE) and the United States Environmental Agency (EPA).

Therefore, computers and devices that meet energy star guidelines display an energy star label in all its products (Roy and Bag, 2009).

Finally, the paper is organized as follows: The Section-2 discusses the concept and classification of green business. Section-3 discusses the drivers of the green business and the paper is concluded in Section- 4.

## 2. Green Business

Green business is a relatively new and not well defined word, which can be interpreted in different ways by different people and organizations. What is considered as green by people/organization is differ to others. Furthermore, the definition of green business is becoming undermined by a proliferation of green labeling and standards. These standards are leading some consumers to consider "green labels" to simplify a marketing tool with little significance behind it. The basic concept of a green business lies in business sustainability. This can be well understood by both consumers and organizations. But, there is a difference in its implementation to what extent it can be applied in practice. In particular, the business decisions should adopt green are based on good business sense.

### 2.1 Classification of Green Business

Basically, there are five key steps in each business life cycle i.e inputs, process, outputs, environment externalities and marketing (Briyan Titley, 2008), ). To be successfully green, business need not only implement cleaner business practices, but also have better communications with their customers in order to establish their brand and capture the market share for green products. Thus, the strategy of the firm is therefore based not only on the concept of productivity but also on the assessment of the life cycle of products and services. Basing on the five steps of the business life cycle, the Figure -1 shows the two criteria for each steps to adopt the green business practices.

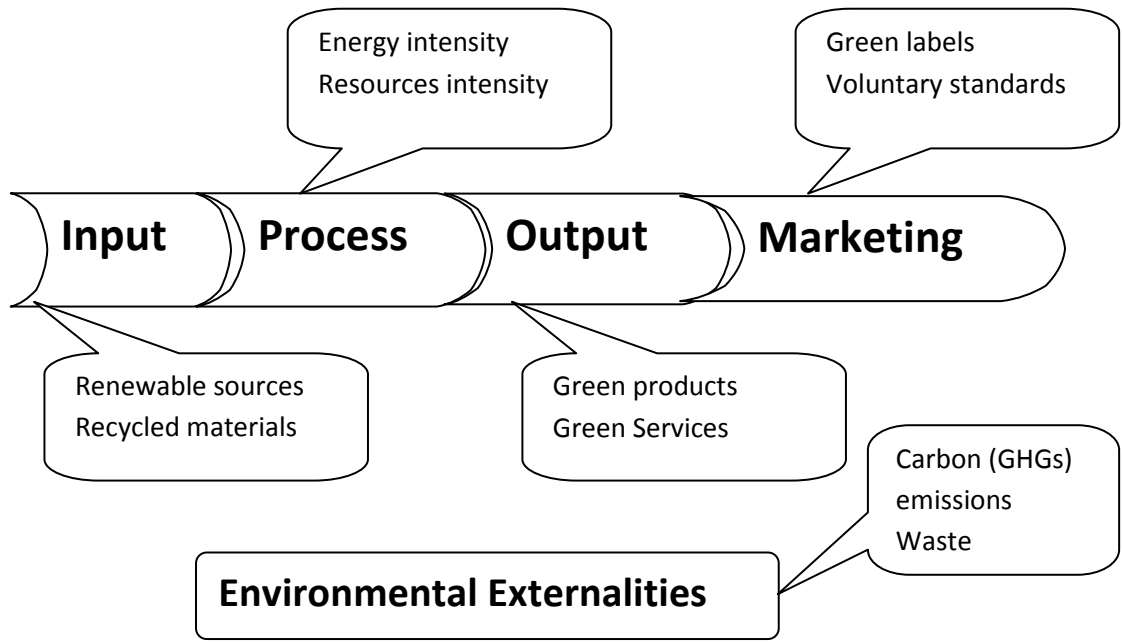

Figure -1: Green measures in business life cycle

Using the green business typology and applying the necessary subjective judgment around some of the criteria identified in the Figure 1, one can lead to differentiate degrees of greenness to different business as follows (Figure -2).

C1: Firms whose activity is to produce environmental goods and services
C2: Firms which have taken active and identifiable steps to change their products and/or process to take substantiality agenda into account
C3: All other firms which have taken some steps to improve then process efficiency or change in their brand image.

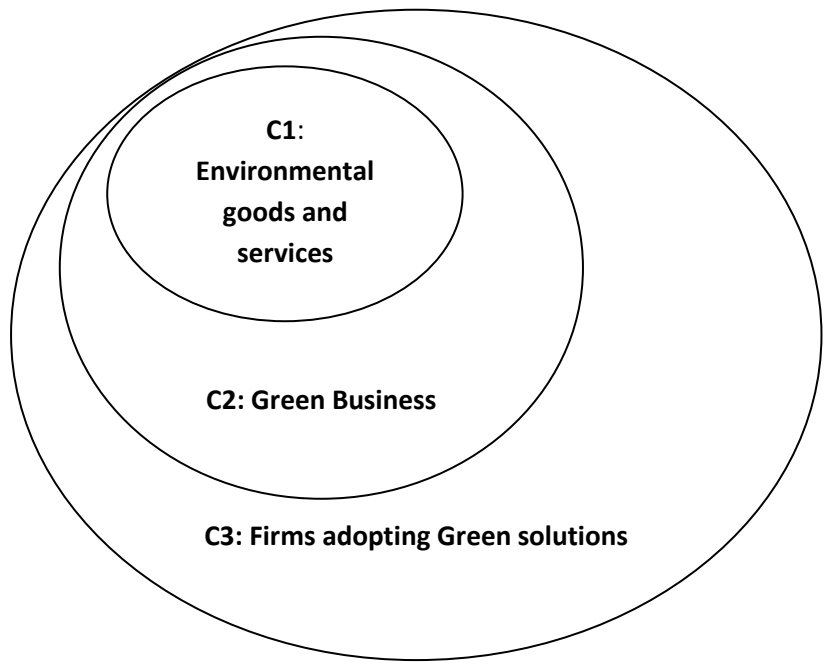

Figure -2: Green business classification

However, it is important to identify categories and measures the greenness of business if it is to be encouraged and promoted.

### 3. Drivers of Green business

The recent expansion of green tech industry and the success of the companies operating in the sector have been driven by changed in the market environment like the regulatory regime and costumer's demand. In fact a study on green tech market in USA highlight the changes in public policy and public awareness as the two key reasons are behind the current clean tech boom ( Clean Tech Venture Capital, 2007). Highly competitive markets and rise in energy prices have also contributed to shift the focus towards greener products and processes. However, it is clear that clean tech products have particularly benefited from a shift in consumer preferences towards greener and cleaner products. In the same time there is a simultaneous trend in actual and expected public policy intervention in the environmental regulation. There are also some important drivers for green business in both public and private sectors. Some of these factors are highlighted below.

### 3.1. Global environmental pressures and public awareness

The public has clearly grown increasingly aware of environmental issues. Over the past few years public awareness and concern about global climate change has risen considerably. Private companies followed closely such changes in public attitudes. An instance, in January 2007, 10 major US companies in collaboration with the four US environmental groups planned for swift action on global climate change known as USCAP (the United States Climate Action Partnership (USCAP)) by federal action on carbon regulations. In addition, the proliferation of organizations focusing on going 'carbon neutral' is an indicator those concerns over climate change have firmly taken root in the public at large. Furthermore, companies around the world are realizing that reducing their environmental footprints can also provide benefits for business. A 2006 survey of 150 companies in the US, UK, France, and Germany by AMR Research found that the top environmental concern among the executives in the survey was 'Energy and Emissions Reductions. As companies are increasingly scrutinized on their stewardship of the environment, corporate social responsibility has become a key component of almost every company's business strategy. Even without government regulations, the energy use and greenhouse gas emission have apparently risen to the top of the list of targeted activities of different companies in private or public domain.

### 3.2. Capital markets acceptance

Now, capital market investors understand the level of public interest in climate change issues. They have started to invest in industries that reduce human impacts on the environment. The clean tech market originally consisted mostly of specialist investment firms and people with a strong environmental focus. However, the world's major public and private equity investment entities are committing capital investment to clean tech industrial production. Nevertheless, the shifting in public awareness and desire for more action both at corporate and government levels has not completely changed the consumer's preferences. Business leaders still believe that the majority of customers will not pay a premium purely for a greener product. Customers want 'green' as an added benefit, whilst they will still consider price, convenience, and performance instead of green attributes. Being green is a benefit which is growing in importance but a majority of customers is still not likely to compromise. However, there is a large and growing segment of the population those prioritizing green factors provide an attractive market for a number of companies. Such niche markets are growing and therefore the opportunities for business are still considerable. These niche green markets will account for several billion in annual revenue. Companies therefore will try to capture the market share for expanding niche of 'green' products by promoting the different qualities of their products- as part of their business strategy. They are also trying to differentiate themselves through 'green branding'. In fact, to be successfully green, the businesses need not only to implement cleaner business practices and reduce their carbon footprint, but also develop better communications with their customers. This will provide the opportunity to establish their brand and create large market demand for green products.

### 3.3. Ensuring market demand for clean tech products or services

Government can directly stimulate market demand by leveraging their own buying power through procurement policies and by making large clean tech purchases. On the other hand government not only increase the market size for such products (which helps bring their prices down through economies of scale), but also set a strong example for ordinary consumers that clean tech purchases are good for the society. Government can also create demand indirectly by requiring a certain amount of energy to be produced from a particular technology, thus incentivizing investment in its production. The most common types of indirect policies in this category are renewable fuels or electricity standards and obligation. Legislation of this type has been in place in many different countries like the Renewable Obligation in the UK and Portfolio Standards in the US. In US, at least 23 different States and the District of Columbia have some form of RES requiring that a certain amount of its electricity usage come from renewable sources.

### 3.4. Creating environmentally-friendly market

One of the most commonly cited proposals for dealing with climate change is establishing carbon price through an emissions 'cap-and-trade' system, where greenhouse gas (or carbon) emissions would be 'capped' at a given level for different companies. Those companies who exceed their allocated limit are required to buy credits to cover their surplus from those companies who emit less than their limit. The world's largest carbon emission cap-and-trade system is the European Emission Trading Scheme which began operational in 2005. In US, a small number of States and other independent bodies have banded together to create emissions markets by placing an actual value on greenhouse gas emissions for the first time. In 2005, the governors of seven States from the Northeast and Mid-Atlantic regions (Connecticut, Delaware, Maine, New Hampshire, New Jersey, New York, and Vermont) established the Regional Greenhouse Gas Initiative (RGGI). In RGGI, the country's first mandatory cap and- trade program designed to reduce the region's greenhouse gas emissions by 10 percent by 2019. A Federal Reserve Bank of Boston analyzes the effects of RGGI and concluded that the program, when coupled with an energy efficiency program, will have a "modest positive impact on gross regional product, personal income, and employment". In particular, RGGI is likely to accelerate growth for some clean tech companies in the region. At the same time a study by the Energy Information Administration (EIA) put the cost of a cap and trade system at federal level to regulate greenhouse gas emissions at $292 over the 2009-2030 time period at the cost of $232 billion (0.10 per cent of GDP) (EIA , 2007).

### 3.5. Providing extra financial backing to clean tech companies

Policies can take the form of subsidies and incentives or tax credits for clean tech products, or taxes on non-clean tech products. These programs are typically financed by taxes and also demonstrated the ability to generate a positive return, which could ultimately lower the customer's bills. A study by the RAND Corporation based on California's energy efficiency program showed an increase in the state's economy of $875 to $1,300 per capita between 1977 and 2000 (Bernstein M, 2000). This shows there is a 40 percent decrease in air pollution emissions from stationary sources and a reduced energy burden on low-income households.

Taxes and tariffs have also had considerable implications for the level of investment in clean technology products. There is strong industry consensus that the bio-fuels boom of 2005-2006 in the US was aided considerably by the federal Volumetric Ethanol Excise Tax Credit (VEETC). Other indirect ways to provide financial backing is through public investment or loan guarantees. It is discussed previously about the important role of public investment in supporting the early stage development of clean tech products, particularly in stimulating the growth of innovative young companies. Analysis from Library House based on Clean Tech
Network data suggests that one of reasons for the success of US and UK in stimulating venture capital investment in clean tech is the participation of public investment in supporting early stage development of small and medium companies. In general, there appears to be a positive relationship between the level of public sector engagement with early stage clean tech companies and the number of these which ultimately receive venture capital backing. The public sector has a variety of other tools at its disposal to boost the clean tech industry, like:

i. **Public education investment** - One of the major reasons commonly cited for the emergence of California's Silicon Valley as a major hub of the Clean Tech industry is the presence of two major universities - University of California of Berkeley and Stanford-with world-class scientific research programmes and top business schools. These two institutions graduate a pool of first-class technical researchers and business-savvy students many of whom become entrepreneurs.

ii. **Clean tech incubators and business assistance** - Incubators help young companies to develop the business skills and acumen critical to become the commercially successful. Typical incubators enable their companies to share office space, basic business services, technical support, and equipment in order to save costs. They also generally offer management advice, technical assistance, networking opportunities, consulting services, and obtaining financial assistance. Incubators can be targeted to specific industries, like clean technologies, or open to a broader range of companies, but whatever their form, they are likely to improve the survival rate of new start-ups and speed the product development and commercialization process.

iii. **Public leadership** - Over the past few years, a variety of public and private citizens have all used their public prominence to raise awareness about global climate change. By voicing their backing for 'green business' and clean tech industry development, public leaders can continue to raise awareness on the subject, and send a message to companies investing in clean technologies that they will receive strong public support in that area.

International Journal of Managing Information Technology (IJMIT) Vol.2, No.3, August 2010### 4. Conclusion:

Green business opportunities provide consumers with ecologically sound products and services. These environmental friendly businesses also provide a competitive niche. While green business is not new, this industry is one of the youngest and as such competition may not be as stiff compared with other small business opportunities. As business continues to do more with less, green business opportunities are certainly here for the long haul. In addition to protect and preserve the environment, "going green" is the only way to grow the business by reducing the eliminate waste

The green business landscape is healthy and continues to grow. Green business opportunities exist today for both residential and business customers. Positioning yourself as the leader in a profitable niche today requires to ensure the long-term viability and growth of your green business.

## References:

[1]  Brown, D., J. Dillard and R.S. Marshall. (2006) Triple Bottom Line: A business metaphor
      for a social construct.  www.recercat.net/bitstream/2072/222311/UABDT06-
      2.pdf [ Last visited on May 2010]
[2]  Round Rock (2009), Dell Ranks No. 1 in First TBR Sustainability Index Benchmark Report ,
      http://content.dell.com/us/en/corp/d/press-releases/2009-05-20-TBR-Green-Report.aspx,
      [Last visited on May, 2010]
 [3] Mike Ebbers,Alvin Galea,Michael Schaefer, Marc Tu Duy Khiem (2009),  The Green Data
      Center:           Steps            for           the           Journey           , http://www.redbooks.ibm.com/abstracts/redp4413.html
      [Last visited on May, 2010].
[4] Rakesh Kumar (2007)   "Eight Critical Forces Shape Enterprise Data Center Strategies"
      Gartner,   Inc,       http://www.gartner.com/DisplayDocument?doc_cd=144650&ref= g_fromdoc
      [last visited May, 2010].
[5] MTB (2009) , Green Computing Sun helping partners offer eco friendly services,
      http://www.mbtmag.com/article/194428-Green_computing_Sun_helping_partners
      offer_eco friendly_ services_.php,[Last visited on May, 2010]
[6]  Greg Horn (2006), Living Green: A Practical Guide to Simple Sustainability (Paperback),
      Freedom Press, USA.
[7] S Roy and M Bag (2009),  Green Computing - New Horizon of Energy Efficiency and E30